# AUTOMATED RF PHASE ADJUSTMENT FOR BEAM STABILIZATION IN THE FERMILAB LINAC*

R. R. Chichili[†,1], J. A. Sulskis[1,2], R. Sharankova[3], B. Vamanan[1], S. Ravi[1]  
[1]University of Illinois at Chicago, Chicago, IL, USA  
[2]also at Georgia Tech Research Institute, Atlanta, GA, USA  
[3]Fermi National Accelerator Laboratory, Batavia, IL, USA

## Abstract

The Fermilab Linac experiences longitudinal beam phase drift, leading to increased particle loss, conventionally corrected through labor-intensive manual RF adjustments. This project explores machine learning-based automation for drift correction, employing a prototype-based classification approach. Our model utilizes a 34-dimensional feature set (RF settings and BPM readings) and leverages a 7x27 response matrix for system modeling. To overcome limited real-world data, we generate synthetic data, enhancing model training and generalizability. Custom loss functions, including a surrogate energy-consistent loss and a temporal smoothness constraint, ensure physically plausible drift predictions. The goal is a robust system for autonomous phase adjustments, ensuring stable beam acceleration and reduced manual intervention.

## INTRODUCTION

The Fermilab Linac delivers 400 MeV H⁻ beam. An ion source, a LEBT, a Radio Frequency Quadrupole (RFQ), and a MEBT including a Buncher cavity inject 750 keV beam into the Linac. It consists of 5 drift tube cavities, two side-coupled bunchers, and 7 side-coupled cavities. Environmental changes and instabilities from the ion source and RFQ result in longitudinal beam phase drift, which causes increased particle loss due to emittance blow-up, and is traditionally corrected manually by small adjustments to cavity phase settings [1]. This labor-intensive manual process is a hindrance to maintaining stable beam acceleration.

To solve this, we're exploring a machine learning-based approach. Our model is a prototype-based classifier [2] that learns to map BPM beam phase readings to optimal RF cavity settings. It uses a 34-dimensional feature set (comprising 7 RF cavity phase settings for the RFQ, Buncher and drift tube cavities and 27 Beam Position Monitor (BPM) readings of beam phase). A key part of our strategy is generating synthetic data from this matrix to overcome the scarcity of real-world data and improve the model's ability to generalize.

The model uses custom loss functions that are grounded in physics. A surrogate energy-consistent loss penalizes corrections that lead to implausible energy changes, and a temporal smoothness constraint prevents abrupt adjustments. These functions ensure that the predictions are physically sound. Our overall goal is to build a robust, autonomous system that provides precise phase adjustments, leading to stable beam operation and minimal manual intervention.

## SYSTEM DESIGN

Our system is built around a machine learning-based approach for automated RF phase adjustment in the Fermilab Linac. At its core is a prototype-based classification model [2] designed to identify optimal correction strategies.

### Data Representation and Feature Engineering

The model's input is a 34-dimensional feature set comprising 7 RF cavity settings and 27 BPM phase readings. BPMs measure beam phase relative to the cavity RF reference signal. To model the linear relationship between these parameters, we construct a 7×27 response matrix [3]. This matrix is derived from experimental data using Fast Fourier Transform (FFT) filtering and normalization, providing the fundamental physical relationship $\Delta \mathbf{BPM} = \mathbf{R}^\top \cdot \Delta \mathbf{RF}$ and serving as a basis for synthetic data generation.

### Synthetic Data Generation

Due to the limited availability of real-world experimental data, we generate synthetic data by leveraging the constructed response matrix. This process simulates diverse operational drift scenarios around nominal setpoints, creating a robust training dataset that enhances the model's ability to generalize across various conditions.

### Model Architecture and Classification

We employ a feedforward neural network that maps 27-dimensional BPM input vectors to 7-dimensional output vectors representing nominal RF cavity settings. Nominal conditions were defined as cavity phase settings which resulted in good Linac operation conditions and minimal beam loss. This is a prototype-based classification framework [4] where known nominal configurations are treated as distinct class prototypes. During training, a cross-entropy loss [4] function is used, based on the proximity of the model's output to these prototypes, allowing for a scalable and efficient correction strategy.

---

* This work has been authored in part by FermiForward Discovery Group, LLC under Contract No. 89243024CSC000002 with the U.S. Department of Energy, Office of Science, Office of High Energy Physics and work supported by Discovery Partners Institute https://dpi.uillinois.edu/
† rchic@uic.edu  Code: `https://github.com/jaysulk/uic_fnal_linac_phase_comp`

*Loss Functions*

The model is trained using a composite loss function that goes beyond standard regression and classification metrics. It incorporates a Surrogate Energy-Consistent Loss to ensure physically plausible corrections, a Temporal Smoothness Constraint to prevent abrupt adjustments, and a Fit-Loss to guarantee that the model's output, when applied through the response matrix, accurately reconstructs the observed BPM readings. The total loss is a weighted sum of these components, allowing for nuanced optimization.

## IMPLEMENTATION

*Synthetic Data Generation and Augmentation*

To overcome data scarcity, we use a physics-based, five-step workflow to generate synthetic data that preserves beam dynamics.

**1. Response Matrix Construction:** To extract the longitudinal response matrix, we follow the method described in [3] and record BPM phase measurements while oscillating the cavity phase settings of the 7 cavities at the same time at different frequencies. Then we build a 7×27 matrix **R** from experimental cavity phase scan data, where each row shows the response of all 27 BPMs to a 1° phase change in a single cavity [3]. This matrix captures the linear relationship:

$$\Delta \mathbf{BPM} = \mathbf{R}^\top \cdot \Delta \mathbf{RF}$$

**2. Drift Simulation:** To simulate cavity phase drift, we apply cavity-specific perturbations to the response matrix. The drift of the BPMs is calculated as

$$\mathbf{BPM}_{\text{drift}} = \mathbf{R}^\top \cdot \begin{bmatrix} s_1 \\ \vdots \\ s_7 \end{bmatrix},$$

where for a single cavity $k$, its drift magnitude is $s_k \in \{0.1, 0.2, 0.3, 0.4, 0.5\}°$, while all other cavities have $s_j = 1$ (no drift). This process generates 35 distinct scenarios (7 cavities × 5 magnitudes).

**3. Iterative Correction Process:** We simulate iterative correction. The process initializes with the drifted BPM readings and repeats until the RMS (Root Mean Square) error between residual and corrected BPMs is less than 0.1°. In each iteration, a neural network predicts a correction, noise is added, and the residual is updated:

1. **Initialize**: $\mathbf{BPM}_{\text{res}} \leftarrow \mathbf{BPM}_{\text{drift}}$

2. **Repeat until convergence** (RMS < 0.1°):

    (a) Predict: $\Delta \mathbf{RF}_{\text{pred}} = \text{NN}(\mathbf{BPM}_{\text{res}})$

    (b) Add noise: $\Delta \mathbf{RF}_{\text{noisy}} = \Delta \mathbf{RF}_{\text{pred}} + \mathcal{N}(0, \text{RMS}_{\text{prev}})$

    (c) Compute correction: $\mathbf{BPM}_{\text{corr}} = \mathbf{R}^\top \cdot \Delta \mathbf{RF}_{\text{noisy}}$

    (d) Compute RMS between $\mathbf{BPM}_{\text{res}}$ and $\mathbf{BPM}_{\text{corr}}$

    (e) Update: $\mathbf{BPM}_{\text{res}} \leftarrow \mathbf{BPM}_{\text{corr}}$

**4. Correction Extraction:** The total correction, $\Delta \mathbf{RF}_{\text{final}}$, is the last noisy prediction ($\Delta \mathbf{RF}_{\text{noisy}}$) before the convergence criterion is met.

**5. Synthetic Sample Creation:** Each training sample consists of a drifted BPM reading ($\mathbf{BPM}_{\text{drift}}$) as input and the final correction ($\Delta \mathbf{RF}_{\text{final}}$) as the target.

This process ensures that the generated synthetic data adheres to real accelerator physics principles, with additive beam dynamics where $\text{RF}_{\text{synth}} = \text{RF}_{\text{real}} + \Delta \mathbf{RF}_{\text{final}}$ and $\text{BPM}_{\text{synth}} = \text{BPM}_{\text{real}} + \mathbf{BPM}_{\text{corr}}$. This efficient scaling approach allows a **n** number of real-world data points to generate **35×n** of physics-consistent synthetic samples, creating a robust training corpus to handle Linac drift scenarios.

*Neural Network Architecture*

**Model Based Drift Correction** As hinted above in the response matrix construction process, one way is to find new cavity settings using linear correction approach [1]. Computationally, this involves iteratively solving a linear system of equations with a nominal trajectory to obtain the best cavity setting using the (pseudo-inverse of) response matrix. Our goal is to avoid the iterative procedure and directly obtain the cavity settings using **R** using a neural network model.

As illustrated in Fig. 1, our neural network processes a 27-dimensional input vector of BPM readings through four fully connected hidden layers (256, 128, 64, and 32 neurons respectively), all utilizing ReLU activation functions for nonlinear feature extraction. This hierarchical structure progressively refines the signal representation before culminating in a 7-neuron output layer that predicts RF cavity adjustments. Later, the network implicitly utilizes a conceptual "prototype" layer, where its output is compared to predefined nominal cavity configurations to guide the classification and regression tasks.

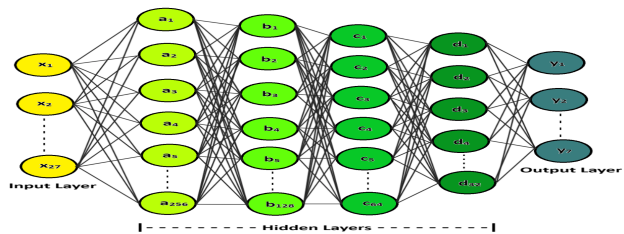

Figure 1: Model architecture.

*Training Process*

After standardization of data, the model is trained and optimized using mini-batch gradient descent with the Adam optimizer. The objective here is to minimize a composite loss function, $\mathcal{L}_{\text{total}}$, which is a weighted sum of five key components:

$$\mathcal{L}_{\text{total}} = w_1 \mathcal{L}_{\text{reg}} + w_2 \mathcal{L}_{\text{cls}} + w_3 \mathcal{L}_{\text{energy}} + w_4 \mathcal{L}_{\text{temp}} + w_5 \mathcal{L}_{\text{fit}}.$$

Here, the components are defined as:

- **Regression Loss ($\mathcal{L}_{\text{reg}}$):** Mean Squared Error (MSE) to ensure numerical accuracy between the predicted ($\hat{\mathbf{y}}$) and nominal ($\mathbf{y}$) cavity settings:

$$\mathcal{L}_{\text{reg}} = \frac{1}{N} \sum_{i=1}^{N} ||\hat{y}_i - y_i||^2.$$

- **Classification Loss ($\mathcal{L}_{\text{cls}}$):** A cross-entropy loss [4] based on the proximity of the output to predefined nominal cavity prototypes ($\mathbf{P}_k$):

$$\mathcal{L}_{\text{cls}} = -\sum_{k} p_k \log(\text{softmax}(-\text{dist}(\hat{y}, P_k))).$$

- **Energy Consistency Loss ($\mathcal{L}_{\text{energy}}$):** A physics-inspired variable penalizing inconsistencies between predicted corrections ($\mathbf{f(x)}$) and estimated beam energy changes:

$$\mathcal{L}_{\text{energy}} = \lambda_e \sum_{i=1}^{N} w_i \, ||f(x_i) - y_i||^2.$$

- **Temporal Smoothness Loss ($\mathcal{L}_{\text{temp}}$):** A regularization variable to promote physically plausible, gradual corrections across timesteps:

$$\mathcal{L}_{\text{temp}} = \lambda_t \sum_{t=1}^{T-1} ||f(x_{t+1}) - f(x_t)||^2.$$

- **Fit-Loss ($\mathcal{L}_{\text{fit}}$):** A variable that ensures the predicted cavity settings, when applied through the response matrix, accurately reconstruct the observed BPM readings($\mathbf{x}$).

$$\mathcal{L}_{\text{fit}} = \frac{1}{N} \sum_{i=1}^{N} ||\mathbf{R}^\top \cdot \hat{y}_i - x_i||^2.$$

*Testing and Evaluation Protocol*

Our comprehensive evaluation protocol assesses the model's performance on an unseen test dataset. We benchmark our model against baseline method, such as linear regression in Ref. [3]. Model efficacy is demonstrated through visual plots, including loss curves (where train and test losses are monitored to assess model's convergence and prevent overfitting) and trajectory fit plots that compare the observed BPM trajectories against those reconstructed from the model's predicted corrections. This protocol ensures the system is not only computationally accurate but also physically consistent and robust.

## RESULTS

Our model's performance was evaluated by its ability to correct longitudinal beam drift. The model's efficacy is determined by comparing the predicted BPM trajectory (derived by applying the predicted RF cavity adjustments through the Linac longitudinal response matrix) to the actual nominal beam trajectory.

The training process's effectiveness is shown by the model's loss curves. We found that these curves track the total loss for both training and test datasets across epochs, with both consistently decreasing. The close convergence of the test and training loss curves demonstrates that the model learned the data's underlying patterns without overfitting.

As an illustration of the model's ability to correct drift, Fig. 2 presents a representative plot from our standardized test dataset. This visualization directly compares the nominal BPM trajectory (actual/non-drifted, shown in blue) against the BPM trajectory that is reconstructed by applying the model's predicted cavity settings (generated on drifted data as its input, shown in orange). This comparison visually demonstrates how the model's adjustments aim to steer the beam from its drifted state to its actual nominal, non-drifted state. We found that the prediction error between predicted and actual trajectories are about 5% showing that the model fits reasonably well over the full range of BPMs.

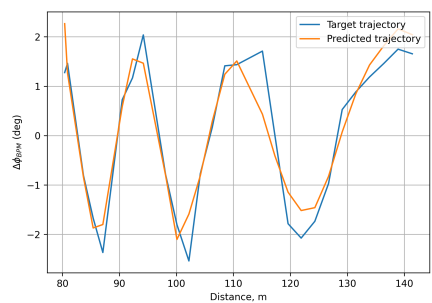

Figure 2: Trajectory fit check on test sample.

## DISCUSSION AND FUTURE WORK

A key challenge is the evolving nature of beam drift, which currently requires periodic model retraining. Future work will explore self-adaptive models using online meta-learning for continuous adjustment. This would allow the system to "learn while correcting", ensuring long-term accuracy without manual updates.

## CONCLUSION

This work successfully developed and validated a machine learning system for automated longitudinal beam drift correction in the Fermilab Linac. By using a physics-informed synthetic data strategy, the model overcomes the challenge of limited real-world data. The custom neural network, with its comprehensive composite loss function (including regression, classification, energy consistency, temporal smoothness, and fit-loss), ensures that the predicted RF cavity adjustments are both accurate and physically consistent. This innovative approach demonstrates promising initial results, establishing a strong foundation for future real-time testing and model refinements, ultimately contributing to more efficient and reliable particle accelerator operations at Fermilab and potentially similar facilities worldwide.